\documentclass[aoas,nameyear,dvips]{arximspdf}

\doi{10.1214/09-AOAS259}
\volume{3}
\issue{4}
\pubyear{2009}
\firstpage{1493}
\lastpage{1504}

\makeatletter
\newtheorem{Theorem}{Theorem}

\newproclaim{example}{Example}
\makeatother

\begin{document}
\begin{frontmatter}

\title{Incompatibility of trends in multi-year estimates
from the American Community Survey}
\runtitle{Trends for ACS}

\begin{aug}
\author{\fnms{Tucker} \snm{McElroy} \ead[label=e1]{tucker.s.mcelroy@census.gov}\corref{}}
\runauthor{T. McElroy}
\affiliation{U.S. Census Bureau}
\address{Statistical Research Division\\ U.S. Census Bureau \\ 4600
Silver Hill
Road\\
Washington, DC 20233-9100\\
USA} 
\end{aug}

\received{\smonth{11} \syear{2008}}
\revised{\smonth{3} \syear{2009}}

%
\begin{abstract}
The American Community Survey (ACS) provides one-year (1y), three-year (3y)
and five-year (5y) multi-year estimates (MYEs) of various demographic
and economic
variables for each ``community,'' although the 1y and 3y may not be
available for communities with a small population. These survey
estimates are not
truly measuring the same quantities, since they each cover
different time spans. Using some simplistic models, we demonstrate
that comparing different period-length MYEs results in spurious
conclusions about trend movements. A simple method utilizing
weighted averages is presented that
reduces the bias inherent in comparing trends of different MYEs.
These weighted averages are nonparametric, require only a short span
of data, and are
designed to preserve polynomial characteristics of the
time series that are relevant for trends. The basic method, which only requires
polynomial algebra, is outlined and applied to ACS data. In some
cases there is an improvement to comparability, although a final
verdict must await additional ACS data. We draw
the conclusion that MYE data is not comparable across different
periods.
\end{abstract}

%
\freefootnotetext[]{\textit{Disclaimer}.
This paper is released to inform interested parties of ongoing
research and to encourage discussion of work in progress. Any views
expressed are those of the author and not necessarily those of the
U.S. Census Bureau.}
\end{frontmatter}
%
\section{Introduction}\label{sec1}
The American Community Survey (ACS) replaces the former Census Long
Form, providing timely estimates available throughout the decade.
The ACS sample size is comparable to that of the Census Long Form;
variability in the sampling error component of the ACS is
partially reduced through a rolling sample [Kish (\citeyear{Kish81})]. The
rolling sample refers to the pooling of sample respondents over
time---in some cases this may be viewed as an approximate temporal
moving average of single period estimates. In particular,
estimates from regions with at least $65\mbox{,}000$ people are produced
with a single year of data, whereas if the population is between
$20\mbox{,}000$ and $65\mbox{,}000,$ then three years of data are combined, and if the
population is less than $20\mbox{,}000,$ then five years of data are
pooled. A somewhat dated
overview of the ACS can be found in Alexander (\citeyear{Alexander98}). More current
details can be found in the Census Bureau (\citeyear{Census06}) and Torrieri (\citeyear{Torrieri07}).

In order to examine longer time series of ACS data, it is necessary
to examine older estimates published for a small group of regions
in the Multi-Year Estimates Study (MYES), which is publicly available
at
\href{http://www.census.gov/acs/www/AdvMeth/Multi\_Year\_Estimates/online\_data\_year.html}
{www.census.}\break
\href{http://www.census.gov/acs/www/AdvMeth/Multi\_Year\_Estimates/online\_data\_year.html}
{gov/acs/www/AdvMeth/Multi\_Year\_Estimates/online\_data\_year.html}.

The MYES was a trial study for the ACS that produced one, three and
five year estimates for counties
included in the 1999--2001 demonstration period and their
constituent geographies, using data from $1999$ through $2005$.
The Multi-Year Estimates (MYEs) are divided according to period-length---either
one-year (1y),\footnote{Technically, the 1y are not MYEs, but we will
ignore this for didactic purposes.} three-year (3y) or five-year
(5y)---the time period, the county and the geographic type within the
county (e.g., school district). There are hundreds of variables
available, which are broken into four categories: demographic,
economic, social and
housing. Most of the variables are totals, averages, medians or
percentiles.

Because some counties have a low population, it was deemed desirable
by the~U.S. Census Bureau to decrease sampling error for smaller
geographies and subpopulations by using a rolling
sample; a discussion of issues associated with this methodology can be
found in
the National Academy of Sciences Panel on the Functionality and
Usability of Data from the American Community Survey [Citro and Kalton
(\citeyear{Citro07})]. In essence, responses over a 3y or even a 5y span
are gathered together into one database, and a statistic of
interest is computed over the temporally enlarged sample. In many
cases, this is approximately equal to computing a~simple moving
average of 1\hspace*{-1pt}y estimates. This is known as a rolling sample---see Kish
(\citeyear{Kish81,Kish98}) and Alexander (\citeyear{Alexander01}) for a discussion.
For larger counties, the~1y MYE would be available
as well. The question of whether each year should
be equally weighted was addressed in Bell (\citeyear{Bell98}) and Breidt (\citeyear{Breidt07});
since all the responses are pooled
in the 3y and 5y cases, the U.S. Census Bureau judged that it would be
impractical to use some
alternative weighting scheme (such as weighting the most recent
year of data more highly). Hence, the MYEs are formed from
contributions over multiple years that are \textit{equally} weighted.
Although this approach is simple, one repercussion
is that some lag (or time delay) is induced by the use of rolling
samples (whereas an unequal weighting scheme can be devised such
that time delay is reduced or eliminated for certain components of
the time series).

The time delay effect is easy to understand in the case that the
data is a simple polynomial, such as a line or a quadratic. In the
former case, a three-period average induces a time delay
of exactly one time unit, whereas the five-period average
delays the line by two time units. For higher degree polynomials
the delay is not exact, and yet visually there is a definite shift
in the graph of one or two units. Assuming that trends in ACS MYEs
are locally given by low-degree polynomials, this brief discussion
illustrates the problem with comparing MYEs of different period
lengths (and this is further expounded in Sections \ref{sec2} and \ref{sec3} below).
In particular, making comparisons across
regions of MYEs of different period lengths will in general lead
to false conclusions and spurious deductions, and therefore should
be avoided. This paper assesses the extent of this problem
through some extremely simple models, and proposes a class of
trend-preserving weighted averages that can be used to illustrate
and identify the sorts of false conclusions arising from such
inter-period comparisons. The perspective of this author is that
such cross-period MYE comparisons should not be made for reasons
discussed in the subsequent sections. Although use of the
proposed weighted averages in this paper may well, in some cases,
reduce the quantity of spurious
conclusions drawn from the data, it is acknowledged that they do
not provide a full solution to the problem of incomparability.

In Section \ref{sec2} we provide additional discussion of the construction
of MYEs, explicating the practical factors militating against
inter-period comparisons. Then in Section \ref{sec3} we discuss a simple model
for MYEs that focuses on the temporal aspects, while ignoring
sampling error for simplicity. Using this formal approach, we can
illustrate in a quantitative fashion the pitfalls that may occur from making
cross-period MYE comparisons. In Section \ref{sec4} we propose a system of
weighted averages that preserve any local polynomial trends,
ensuring that these trends for~1y,~3y and 5y are identical after
application of the weights. This is a general technique based on
simple time series analysis and polynomial algebra, and we apply
it in the linear trend case to MYE data in Section \ref{sec5}, making
use of the newly available ACS data extended by the trial period of
the MYES.
Through several examples, we illustrate the dangers of making
inappropriate comparisons, that is, cross-region comparisons
involving MYEs of different period lengths. Finally, Section \ref{sec6}
summarizes the results of the paper and the main difficulties in
inter-period comparisons.
%
\section{Practical issues in making comparisons}\label{sec2}
Beyond the issues of time delay raised in the Introduction and
further described below, there is a problem comparing MYEs of
different period lengths due to the differences in how the
estimates are constructed. A detailed discussion of these issues
is beyond the scope of this paper [for more information the reader is
referred to
Fay (\citeyear{Fay07}), Starsinic and Tersine (\citeyear{Starsinic07}), and Tersine and Asiala
(\citeyear{Tersine07})], but here we briefly
highlight some relevant points.

In the construction of MYEs a weighting method is used that is
different for~1y versus 3y and 5y. In the former case, baseweights
are used that are defined as the inverse of sampling probabilities,
with some differences between Housing Units~(HU) and Group Quarters (GQ).
Next, there is a nonresponse adjustment followed by the application of
controls to a set of independent HU estimates derived from the U.S.
Census Bureau's Population Estimates Program (GQs are handled with
separate controls).
For the 3y and 5y estimates, similar weighting and adjustments are
made, but based off of data pooled over the whole three years and
five years respectively. Moreover, housing unit controls are
further modified by the so-called g-weighting (a type of calibration)
[see Fay (\citeyear{Fay05,Fay06,Fay07})], with the objective of reducing (sampling error) variances
at the sub-county aggregation level. This process involves linking
administrative records data with the~ACS sampling frame [Starsinic
and Tersine (\citeyear{Fay07})].

As a result of g-weighting, the 3y and 5y estimates are fundamentally
different in
their construction from the 1y. We also point out that, apart from
the g-weighting, there is also the issue of additional pooling in
3y and 5y prior to weighting and nonresponse adjustment; thus, a 5y
estimate will have effectively five times as many sample cases
receiving weighting over the 1y estimate. Furthermore, the
population controls will vary between MYEs, since the vintage of
the population estimates will correspond to the final year in the
particular MYE. So the 3y MYE for 2005, 2006 and 2007 is controlled to the
average population for those years at a 2007 population vintage,
whereas the 1y MYE for each of the corresponding years
2005, 2006 and 2007 will each be based off population vintages
from those three years; this further interferes with comparability.
A related issue is inflation adjustment for monetary variables, which
is handled by
controlling to dollars in the latest year of the period.

These are fundamental incompatibilities; one may see that 1y, 3y
and 5y are really measuring different quantities. The weighted
average methodology of this paper---presented below---can address
the issue of pooling in an approximate fashion, but does not
provide a resolution to the effects of g-weighting, nonresponse
adjustment and variable (population and monetary) vintages.
However, given that it is common in trend analysis of demographic
and economic time series to compare data that have no common basis
of measurement [e.g., consumption versus income is analyzed for
co-integration in Engle and Granger (\citeyear{Engle87})], it is only vital to
account for
time delay shifts in the respective time series. Although such
weighted MYEs are not strictly comparable, they can still be used as
subjects in such a longitudinal or multivariate analysis, just as
similar situations are treated throughout the social sciences [see
Granger (\citeyear{Granger04})].
%
\section{Comparing MYEs}\label{sec3}
This section develops the issue of comparability in a~mathematical
framework, so that we can obtain a quantitative view of why
inter-period comparisons are problematic.
The MYEs are currently available as an annual time series, and we
use the notation $Y^{(k)}_t$ for the $k$y MYE available at year~$t$, where $k = 1, 3, 5$. We define the Simple Moving Average (SMA)
polynomial of order $k$ by
\[
\Theta^{(k)} (z) = \frac{1}{ k} ( 1 + z + \cdots+
z^{k-1}).
\]
As usual, $B$ denotes the backshift operator. Because of the
method of construction of the MYEs described in Section \ref{sec1}, we
might think that $Y^{(5)}_t = \Theta^{(5)} (B) Y^{(1)}_t$ and
$Y^{(3)}_t = \Theta^{(3)}
(B) Y^{(1)}_t $ are approximately true equations [such an
assumption is used for certain variance calculations in Citro and
Kalton (\citeyear{Citro07})]. However, in our experience this approximation is
poor for many variables, and is fair for only a few
variables---typically those involving linear statistics such as
totals and averages. Therefore, we adopt the following error
model for the purpose of demonstrating issues of comparability of trends:
%
\begin{equation}
\label{eq:model1}
Y^{(k)}_t = \Theta^{(k)} (B) \mu_t + \varepsilon^{(k)}_t,
\end{equation}
for $k = 1,3,5$. Here $\mu_t$ is a common deterministic trend
function, and the errors~$\varepsilon^{(k)}_t$ include sampling error,
serially correlated stochastic trend perturbations and
``nonadditive error,'' that is, the error attributed to assuming a
moving average relationship to be valid. We will not be concerned
with the statistical properties of these errors, though they are
assumed to be identically distributed in $t$ with mean zero.
The common trend $\mu_t$ is conceived of
abstractly, and does not necessarily have a fundamental interpretation
in terms of
the population trend. Although other models could be considered [such as
$Y^{(k)}_t = \Theta^{(k)} (B) ( \mu_t + \varepsilon^{(k)}_t)$],
(\ref{eq:model1}) will be sufficient for our illustrative purposes.

Now suppose that we have two time series of MYEs, denoted
$Y^{(k)}_t$ (with trend~$\mu^Y_t$ and error process
$\varepsilon^{(k)}_t$) and $Z^{(k)}_t$ (with trend $\mu^Z_t$ and error
process $\eta^{(k)}_t$). These MYEs may correspond to two
different geographical regions, and a practitioner may be
interested in comparing the trends $\mu^Y_t$ and $\mu^Z_t$, either
at several time points or perhaps at just one time $t_0$. Formally,
we might consider the following hypotheses, although many others
are conceivable:
\begin{eqnarray*}
 H_0 \dvtx \mu^Y_{t_0} &=& \mu^Z_{t_0}, \\
 H_a \dvtx \mu^Y_{t_0} &>& \mu^Z_{t_0}.
\end{eqnarray*}
In this formulation, the values of the mean at time $t_0$ simply
become parameters, and it is the statistician's task to devise
parameter estimates that are accurate and precise. Since
typically in applications it is desirable to make trend
comparisons in real-time, any estimators must be a function of
present and past data only, that is, $\hat{\mu}^Y_{t_0}$ and
$\hat{\mu}^Z_{t_0}$ are functions of the MYE series at times $t_0,
t_0 -1, \ldots.$ The simplest unbiased estimators are
$\hat{\mu}^Y_{t_0} = Y^{(1)}_{t_0}$ and $\hat{\mu}^Z_{t_0} =
Z^{(1)}_{t_0}$, but the 1y MYEs are not always available.
Suppose that the first region ($Y$) includes 1y, 3y and 5y period
MYEs, but the second ($Z$) includes only 3y and 5y.

Commonly, users of MYEs (despite official cautions to the
contrary) will take $\hat{\mu}^Y_{t_0} = Y^{(1)}_{t_0}$ and
$\hat{\mu}^Z_{t_0} = Z^{(3)}_{t_0}$ [or even equal to
$Z^{(5)}_{t_0}$], even though the latter is a biased estimate [due
to the phase delay of $\Theta^{(3)} (B)$; see below] of the trend. We
refer to
this as the ``inapt'' comparison. Seeking to mitigate the phase
delay, we can put both trend estimates on an equal footing by
taking $\hat{\mu}^Y_{t_0} = Y^{(3)}_{t_0}$ and
$\hat{\mu}^Z_{t_0} = Z^{(3)}_{t_0}$. Now both trend estimates are
biased, but at least they are biased in a similar fashion; this
will be called the ``untimely'' comparison. A ``proper'' comparison
is one in which both estimates are unbiased for their respective
trend values. Of course, even for a~proper comparison Type I and II
errors will occur due to statistical uncertainty, but at least the
bias will be eliminated.

One could test the hypothesis of equal trends via
$\hat{\mu}^Y_{t_0} - \hat{\mu}^Z_{t_0}$; this has the following
expectation for the inapt comparison: $\mu^Y_{t_0} - (\mu^Z_{t_0} +
\mu^Z_{t_0 -1} + \mu^Z_{t_0 -2})/3$, which need not be zero under
$H_0$. For the untimely comparison, the expectation would be
\[
\bigl( ( \mu^Y_{t_0} - \mu^Z_{t_0} ) + ( \mu^Y_{t_0 -1} - \mu^Z_{t_0
-1} )
+ ( \mu^Y_{t_0 - 2} - \mu^Z_{t_0 - 2} ) \bigr) /3.
\]
If the trends agree at times $t_0$, $t_0 -1$, and $t_0 -2$, this
quantity is zero; however, some bias is to be expected under $H_0$.
In contrast, it is clear from the definition of
the proper comparison that the mean of $\hat{\mu}^Y_{t_0} -
\hat{\mu}^Z_{t_0}$ is zero under $H_0$.

From this discussion, we see that making inferences about trends
based on a~direct use (i.e., by looking just at the values rather than
some more
complicated statistics) of MYEs of different period lengths leads
to bias even in the case that a highly idealized model holds true.
The incidence of spurious conclusions (i.e., Type I errors) can
be reduced by making proper comparisons, and we explore this
further in the following section. However, even proper
comparisons have their limitations, and our attitude is that MYEs
of different period length should not be compared; using a proper
comparison provides an improvement, but false conclusions can
still be obtained (not to speak of the practical issues raised in
Section~\ref{sec2}).

We note that the incomparability of trends increases with the
dispersion of the errors $\varepsilon^{(k)}_t$; if these errors were
zero, then the rolling sample would be exactly a moving average,
and a proper comparison would enable full comparability of MYE
trends. A crude assessment of the size of these errors, relative
to the trend, is given by the ``Noise-Signal Ratio'' (NSR)
\[
\frac{ \varepsilon_t^{(k)}}{ \Theta^{(k)} (B) \mu_t } =
\frac{ Y_t^{(k)} }{ \Theta^{(k)} (B) \mu_t } -1.
\]
This is only well-defined when $\Theta^{(k)} (B) \mu_t $ is nonzero,
and we generally suppose that it is positive at all times. Since
we do not know $\mu_t$, we can substitute $Y^{(1)}_t$ when the 1y
MYEs are available. Then for $k=3,5,$ we have $Y^{(k)}_t /
\Theta^{(k)} (B) Y_t^{(1)} - 1$ as our estimate of the NSR.
For convenience, we will instead use logarithms of noise and signal,
which are
approximated (by first-order Taylor series) by the former
expression:
\[
\mathit{NSR}_t^{(k)} = \log Y_t^{(k)} - \log\Theta^{(k)} (B) Y_t^{(1)}
\]
for $k=3,5$. Computing this quantity at all available times $t$, we
define a compatibility measure by
\[
C^{(k)} = \max_t \bigl| \mathit{NSR}_t^{(k)} \bigr|.
\]
If this measure is small, for example, $C^{(k)} = 0.01$, then the
rolling sample is
well-approximated by a moving average, and the proper comparison
is more meaningful.
%
\section{Trend-preserving weighted averages}\label{sec4}
In what follows, the function of the model (\ref{eq:model1}) is to
illustrate the incomparability of MYEs of different period length;
we are not interested in fitting the model to actual MYEs in order
to pursue statistical inference. In this sense, the model only
serves a pedagogical purpose. Next, suppose that $\mu_t$ is given
by a polynomial of degree $d$ in $t$. Is it possible to find sets
of weighted averages, or linear filters, such that when applied to
each MYE the trends will coincide? That is, if we view the
underlying trend of the $k$y MYE as $\Theta^{(k)} (B) \mu_t$, then
we seek three filters $\Psi^{(k)} (B)$ such that $\Psi^{(k)} (B)
\Theta^{(k)} (B) \mu_t$ is the same for each $k=1,3,5$; or, in
other words,
%
\begin{equation}
\label{eq:compatCond}
\Psi^{(1)} (z) = \Psi^{(3)} (z) \Theta^{(3)} (z) = \Psi^{(5)} (z)
\Theta^{(5)} (z).
\end{equation}
Since users are typically interested in comparisons utilizing the
most current data available, it makes sense to formulate our
problem with concurrent filters, that is, filters that only depend on
present and past data. Therefore, each filter is of the form
\[
\Psi^{(k)} (z) = \sum_{j \geq0} \psi_j^{(k)} z^j.
\]
In practice, only a finite number of the coefficients $\psi_j^{(k)}$
are nonzero. Now a filter $\Psi(z)$ will pass (i.e., leave invariant) a
polynomial of degree $d$ if $\Psi(1) = 1$ and
$\frac{\partial^j}{\partial z^j} \Psi(z) \vert_{z=1} = 0$ for $1
\leq j \leq d$ [Brockwell and Davis (\citeyear{Brockwell91}), page 39].
Now using~(\ref{eq:compatCond}) and the fact that $\Theta^{(3)} (z)$ and
$\Theta^{(5)} (z)$ share no common roots, it is easy to see that
\[
\Psi^{(1)} (z) = \Phi(z) \Theta^{(3)} (z) \Theta^{(5)} (z).
\]
We are free to design the polynomial $\Phi(z)$ such that the
polynomial-passing constraints are satisfied; hence, $\Phi(z)$ must
have degree at least $d$. The following theorem describes how to
construct this polynomial.
\begin{Theorem}
\label{thm:trend}
The minimal length concurrent filters $\Psi^{(k)}$ that pass degree
$d$ polynomials and
satisfy (\ref{eq:compatCond}) are given by
\begin{eqnarray*}
\Psi^{(5)} (z) & =& \Phi(z) \Theta^{(3)} (z), \\
\Psi^{(3)} (z) & = &\Phi(z) \Theta^{(5)} (z), \\
\Psi^{(1)} (z) & =& \Phi(z) \Theta^{(3)} (z) \Theta^{(5)} (z),
\end{eqnarray*}
where the coefficients of $\Phi(z)$ are given by the first column
of the inverse of the matrix with entry $jk$ given by
\[
\frac{\partial^{j-1} }{ \partial z^{j-1}} \bigl[ z^{k-1}
\Theta^{(3)} (z) \Theta^{(5)} (z) \bigr] \bigg\vert_{z=1}.
\]
\end{Theorem}
\begin{pf}
Let $\Theta(z) = \Theta^{(3)} (z) \Theta^{(5)} (z)$, with $\phi_k$
the coefficients
of $\Phi(z)$. Applying the polynomial-passing constraints yields
\begin{eqnarray*}
1_{ \{ j = 0 \} } & =& \sum_{l=0}^j {{j}\choose{l}}
\frac{\partial\Phi(z) }{\partial z^l} \bigg\vert_{z=1} \frac{\partial
\Theta
(z)}{\partial z^{j-l} } \bigg\vert_{z=1} \\
& =& \sum_{l=0}^j {{j}\choose{l}}
\sum_{k=0}^d \phi_k \frac{ k!}{ (k-l)!}
\frac{\partial\Theta(z)}{\partial z^{j-l} } \bigg\vert_{z=1} \\
& =& \sum_{k=0}^d \phi_k
\frac{\partial^j }{\partial z^{j} } [ z^k \Theta(z) ] \bigg\vert
_{z = 1}.
\end{eqnarray*}
This is easily rewritten in matrix form, from which the result
follows.
\end{pf}
\begin{example*}[(Linear trends)]
Supposing that the trend is linear and $d=1$, we have
\begin{eqnarray*}
\Psi^{(5)} (z) & =& ( 4 + z + z^2 - 3 z^3)/3, \\
\Psi^{(3)} (z) & =&( 4 + z + z^2 + z^3 + z^4 - 3 z^5
)/5, \\
\Psi^{(1)} (z) & =& ( 4 + 5 z + 6 z^2 + 3 z^3 + 3 z^4 - z^5 - 2
z^6 - 3 z^7 )/15.
\end{eqnarray*}
\end{example*}
\begin{example*}[(Quadratic trends)]
Supposing that the trend is quadratic and $d=2$, we have
\begin{eqnarray*}
\Psi^{(5)} (z) & =& ( 26 -11 z + 3z^2 - 23 z^3 + 14z^4)/9, \\
\Psi^{(3)} (z) & =& ( 26 -11 z + 3z^2 + 3z^3 + 3z^4 - 23 z^5
+ 14z^6 )/15, \\
\Psi^{(1)} (z) & =& ( 26 + 15 z + 18 z^2 -5 z^3 + 9 z^4 - 17z^5 -
6 z^6 - 9 z^7 + 14z^8 )/45.
\end{eqnarray*}
Theorem \ref{thm:trend} has the following interpretation. If one
wishes to make a proper comparison of MYEs (defined in Section \ref{sec3})
that preserves polynomials of order $d$, then the minimal length
linear filters that accomplish this goal are given by Theorem
\ref{thm:trend}.
\end{example*}
%
\section{Illustrations on ACS data}\label{sec5}
We now provide three illustrations of the concepts discussed in
this article. We focus on \textit{Median Household Income} in Pima,
AZ, \textit{Number of Divorced Males} in Lake, IL, and \textit{Median
Age} in Hampden, MA. These three counties are included in the
MYES and, therefore, the data extends back to the year 2000. In
particular, the following MYEs are
available: 2000 through 2007 for 1y, 2001 through 2005 and 2007 for
3y, and 2003 through 2005 for
5y. The year index here refers to the last year that entered into
the sample, and so is consistent with our notation for $Y^{(k)}_t$.
Current ACS estimates are now available for all geographical
regions, covering the 1y years 2006 and 2007, and the 3y MYE 2005--2007
has just become available.
Letting $t$ range between 00 and 05 (referring to the year),
the available database is $Y^{(1)}_{00}, \ldots, Y^{(1)}_{07},
Y^{(3)}_{01}, \ldots, Y^{(3)}_{05}, Y^{(3)}_{07}, Y^{(5)}_{03}, \ldots,
Y^{(5)}_{05}$. In order to apply our methods, we need to impute
(by forecasting) the 3y MYE $Y^{(3)}_{06}$ and
the 5y MYEs $Y^{(5)}_{06}$ and $Y^{(5)}_{07}$. (This is a
provisional necessity, since in the future full time series data
for all counties will be published.)

The missing values are obtained by forecasting them utilizing a
simple random walk model, which is feasible for these time series
based on economic and demographic considerations (to actually fit a
time series model to such a short series is pointless):
\begin{eqnarray*}
\widehat{Y}^{(3)}_{06} & =& \tfrac{1}{2} \bigl(
Y^{(3)}_{05} + Y^{(3)}_{07} \bigr), \\
\widehat{Y}^{(5)}_{06} & =& Y^{(5)}_{05} + \tfrac{1}{2} \bigl(
Y^{(5)}_{05} - Y^{(5)}_{03} \bigr), \\
\widehat{Y}^{(5)}_{07} & =& Y^{(5)}_{05} + \tfrac{2}{2}\bigl(
Y^{(5)}_{05} - Y^{(5)}_{03} \bigr).
\end{eqnarray*}
The MYEs (with imputed values in bold) are given in Table
\ref{table:series1}. The final row of the table gives the various
2007 trend values estimated via the method of Section \ref{sec4} [the data and
calculations are given in McElroy (\citeyear{McElroy09})]. Note that
$Y^{(3)}_{01}$ and $Y^{(5)}_{03}$ are not used in the calculation
of these trend estimates. Although
the Income MYEs follow a linear growth pattern, the Divorce MYEs
fluctuate more in their slope component, whereas the Age MYEs trend
upward very slowly with little noise. Thus, we might say that
Income and Age exhibit linear trend lines, whereas Divorce is
nonlinear; it is important to consider different types of trend
behavior in order to evaluate this paper's method.

\begin{table}[b]
\caption{MYEs for Income, Divorce and Age.
Estimates have been forecast extended for the years $06$
and $07$, written in bold}\label{table:series1}
\begin{tabular*}{\textwidth}{@{\extracolsep{\fill}}lccccccccc@{}}
\hline
&\multicolumn{3}{c}{\textbf{Income MYEs}}
& \multicolumn{3}{c}{\textbf{Divorce MYEs}} & \multicolumn{3}{c@{}}{\textbf{Age MYEs}} \\[-6pt]
&\multicolumn{3}{c}{\hrulefill}
& \multicolumn{3}{c}{\hrulefill} & \multicolumn{3}{c@{}}{\hrulefill} \\
\textbf{Year} & \textbf{1y}& \textbf{3y} & \textbf{5y} & \textbf{1y}& \textbf{3y} & \textbf{5y} & \textbf{1y}& \textbf{3y} & \textbf{5y} \\
\hline
00 & 35223 & & & 14043 & & & 36.40 & & \\
01 & 35615 & 35956 & & 14376 & 14429 & & 37.30 & 36.80 & \\
02 & 37638 & 36780 & & 17866 & 15504 & & 37.00 & 36.80 & \\
03 & 37818 & 37373 & 37510 & 17398 & 16772 & 15473 & 37.10 & 37.00 &
36.70 \\
04 & 38800 & 38739 & 38608 & 15632 & 17156 & 15903 & 37.20 & 37.10 &
36.90 \\
05 & 41521 & 40404 & 40055 & 14591 & 15889 & 15945 & 37.40 & 37.30 &
37.20 \\
06 & 42984 & \textbf{42395} & \textbf{41328} & 20941 & \textbf{17371} & \textbf{
16181} & 37.40 & \textbf{37.35} & \textbf{37.45} \\
07 & 43546 & 44386 & \textbf{42600} & 21844 & 18852 & \textbf{16417} & 37.60
& 37.40 & \textbf{37.70} \\[6pt]
Trend & 43570 & 45223 & 45320 & 19331 & 19217 & 16695 & 37.59 & 37.59
& 38.25 \\
 \hline
\end{tabular*}
\end{table}

As far as the linear approximation to the rolling sample, we can
compute the NSR comparability measure for years 2002--2007 for
$k=3$, and 2004--2007 for $k=5$ (by including the forecasted
data). For Income $C^{(3)} = 0.017$ and $C^{(5)} = 0.020$,
indicating some incompatibility. For the Divorce variable $C^{(3)}
= 0.008$ and $C^{(5)} = 0.042$, indicating a high amount of
incomparability (though most of this comes from the portion of the
data that is forecasted, and thus might be resolved when the real
numbers are published). Finally, the Age variable is highly
compatible with $C^{(3)} = 0.002$ and $C^{(5)} = 0.004$.

Now imagine having two replications of each variable for two separate
regions: county A with all period-length MYEs available, and county
B with a lower population such that only 3y and 5y MYEs are
available. Starting with the Divorce variable, an illustration of the
time delay properties of MYEs
is provided in comparing 1y to one-year-ahead-3y
MYEs; there is a fairly close match up until the 2005 1y MYE and
2006 3y MYE. However, this latter value is imputed, and the true
value could easily have decreased from 2005; instead the imputation
increases merely because there is so much gain in the 2007 3y MYE.
The 2007 ``inapt'' comparison discussed in Section 3 would then
compare 21,844 with 18,852 or 16,417; these are
$-13.7 \%$ and $-24.8 \%$ discrepancies.
If we use weighted averages for comparing trends, the
discrepancies are reduced to $-0.59 \%$ and $-13.6 \%$ respectively
(though given the nonlinear nature
of the trend, we expect the forecasts to be inappropriate, and hence
not as much emphasis should be
placed on the 5y MYEs). In this case the weighted average
methodology helps to properly align the series.

For the Income and Age time series data, which both exhibit linear
trends (with the former having much more variability), the
weighted average method can actually increase discrepancies.
In the former case, the discrepancies of $1.9 \%$ and $-2.2 \%$
become $3.8 \%$ and $4.0 \%$; but for Age the discrepancies of
$-0.53 \%$ and $0.27 \%$ become $0 \%$ and $1.8 \%$ after using
weighted averages. The Age data is very stable, and here an inapt
comparison indicates no change. We have not analyzed these
percentages statistically, as this would require actual modeling of
the time series. Nevertheless, a rough idea about trend
comparability can be deduced by the discussion here.

In summary, we see through these examples that the weighted
average methodology can either increase or decrease discrepancies in some
cases, and seems to work less well with 5y versus 3y MYEs
(although this may also be an artifact of two imputations in the
5y MYEs). Part of this increase in discrepancy is due to the
weighted averages increasing the overall variance (even if they
reduce the bias of direct comparisons, as discussed in Section \ref{sec3});
if in (\ref{eq:model1}) we make the crude assumption that the
errors $\varepsilon_t^{(k)}$ are i.i.d., then the linear weights inflate
the variance by a factor of $1.16$ and $3$ respectively for the 3y
and 5y MYEs. For the 1y MYE the variance is multiplied by $0.48$,
but of course this MYE has the greatest variability since its
sampling error component is largest. This variance inflation can
be corrected by imposing extra conditions on the filter
coefficients, but the result would be an even longer set of
weights. It can also be observed that the random walk model used
for forecasting is poorly suited to the Divorce data, since the
change in direction from 2003 to 2004 in the 1y MYE is not
reflected in the corresponding time-delayed 5y MYEs of
2005--2006. A more definitive study would not rely on
imputations, and would be concerned with the qualitative aspects of
trends produced by weighted averages; such a study must wait at
least five years due to the current ACS publication schedule.
%
\section{Conclusion}\label{sec6}
The aim of this paper is first to discuss the challenges in
comparing cross-period MYEs. Due to the way in which MYEs are
constructed, it is apparent that 1y, 3y and 5y MYEs are different
time series---and not just time-lagged or smoothed versions of
some underlying series; they are estimates of \mbox{different} fundamental
quantities (see Section \ref{sec2}). Nevertheless, this fact does not preclude
a user from
making cross-period comparisons, any more than it would be
forbidden to search for common trends in economic or demographic
data. Therefore, the second aim of this paper is to
quantitatively assess what sorts of mathematical and statistical
problems will arise in such comparisons (see Sections \ref{sec3} and \ref{sec4}).
As a third aim, the weighted averages method can be used to reduce the bias
inherent in such cross-period comparisons [under certain
quasi-linear assumptions such as (\ref{eq:model1})]; even so, the
statistical variation in MYEs is such that sizeable discrepancies
can still crop up, as demonstrated in Section \ref{sec5}.

In summary, the author wishes to echo the strong cautions against
making cross-period comparisons issued by the
U.S. Census Bureau [see Beaghen and Weidman~(\citeyear{Beaghen08}) and Citro and
Kalton (\citeyear{Citro07})].
At this point the weighted average methodology mainly serves to
identify fairly egregious types of false conclusions derived from
such unwarranted comparisons, but perhaps it can also serve as
a~building block for future work on comparability and usability
issues in the ACS.

\section*{Acknowledgments}
This paper was greatly improved by helpful discussions with two
anonymous Referees, the Associate Editor and the Editor, as well as fruitful
discussions with Lynn Weidman and Alfredo Navarro of the U.S.
Census Bureau.

\begin{supplement}
\stitle{Income, Divorce and Age Data with Trend Calculations\\}
\slink[doi]{10.1214/09-AOAS259SUPP}
\sdatatype{.zip}
\slink[url]{http://lib.stat.cmu.edu/aoas/259/supplement.zip}
\sdatatype{.zip}
\sdescription{This file contains the Income,
Divorce and Age data of Table 1 in Excel format. Also provided are
the linear trend weighted averages along with compatibility measures
NSR, encoded as Excel formulas.}
\end{supplement}

%

%

\printaddresses

\end{document}